\documentstyle[12pt]{article}
\newcommand{\nn}{\nonumber \\}
\newcommand{\beq}{\begin{equation}} 
\newcommand{\eeq}{\end{equation}}
\newcommand{\bea}{\begin{eqnarray}}
\newcommand{\eea}{\end{eqnarray}}

\newcommand{\AC}{{\mathcal{A}}}
\newcommand{\BC}{{\mathcal{B}}}
\newcommand{\CC}{{\mathcal{C}}}

\newcommand{\VC}{{\mathcal{V}}}
\newcommand{\MC}{{\mathcal{M}}}

\newcommand{\QC}{{\mathcal{Q}}}

\newcommand{\vh}{{\widehat{\mathcal{V}}}}
\newcommand{\ve}{\varepsilon}
\newcommand{\vf}{\varphi}

\newcommand{\s}{ SO(16)}
\newcommand{\so}{ SO(1,2) \times SO(16)}

\newcommand{\sx}{ SO(1,1) \times SO(16)^\infty}
\newcommand{\sxx}{ SO(16)^\infty}

\newcommand{\su}{ SO(1,3) \times SU(8)}
\newcommand{\ti}{\tau^\infty}

\begin{document}
\renewcommand{\theequation}{\thesection.\arabic{equation}}
\renewcommand{\section}[1]{\addtocounter{section}{1}
\vspace{5mm} \par \noindent
  {\bf \thesection . #1}\setcounter{subsection}{0}
  \par
   \vspace{2mm} } 
\newcommand{\sectionsub}[1]{\addtocounter{section}{1}
\vspace{5mm} \par \noindent
  {\bf \thesection . #1}\setcounter{subsection}{0}\par}
\renewcommand{\subsection}[1]{\addtocounter{subsection}{1}
\vspace{2.5mm}\par\noindent {\em \thesubsection . #1}\par
 \vspace{0.5mm} }
\renewcommand{\thebibliography}[1]{ {\vspace{5mm}\par \noindent{\bf
References}\par \vspace{2mm}}
\list
 {\arabic{enumi}.}{\settowidth\labelwidth{[#1]}\leftmargin\labelwidth
 \advance\leftmargin\labelsep\addtolength{\topsep}{-4em}
 \usecounter{enumi}}
 \def\newblock{\hskip .11em plus .33em minus .07em}
 \sloppy\clubpenalty4000\widowpenalty4000
 \sfcode`\.=1000\relax \setlength{\itemsep}{-0.4em} }
\newcommand\rf[1]{(\ref{#1})}
\def\nn{\nonumber}
\newcommand{\sect}[1]{\setcounter{equation}{0} \section{#1}}
\renewcommand{\theequation}{\thesection .\arabic{equation}}
\newcommand{\ft}[2]{{\textstyle\frac{#1}{#2}}}

\thispagestyle{empty}

\begin{flushright} hep-th/9906106
                   \\ AEI-1999-8
\end{flushright}
\vspace*{1.0cm}

\begin{center}

{\large\bf On Hidden Symmetries in $d=11$ Supergravity\\
and Beyond}\footnote{Invited
talk at the Conference ``Fundamental Interactions: From Symmetries
to Black Holes'', 24 - 27 March, Universit\'e Libre de Bruxelles,
Belgium.}\\

\vspace{1.4cm}

{\sc H. Nicolai}\\

\vspace{1.3cm}

{\em Max-Planck-Institut f\"ur Gravitationsphysik} \\
{\em Am M\"uhlenberg, Haus 5} \\
{\em D-14476 Golm, Germany}\\

\vspace{1.2cm}

\vspace{- 4 mm}  \end{center}
\vfill

\baselineskip18pt
\addtocounter{section}{1}
\par \noindent
  \par
   \vspace{2mm} 

\noindent  

This talk is about hidden symmetries in eleven dimensions, but 
it is equally a tribute to a scientist and friend, who is eminently
visible in four space-time dimensions: Fran{\c c}ois Englert, in whose 
honor this meeting is being held. Therefore, before entering 
{\it dans le vif du sujet} I would like to express my gratitude 
for having had the opportunity and privilege to learn from him and
to work with him, and for all the fun we have had --- involving,
amongst other things, dinosaurs within dinosaurs \cite{englert1}
(the ancestor of all modern inflationary theories) and their 
eleven-dimensional avatars \cite{englert2}, as well as higher states 
of consciousness \cite{englert3} and monster strings \cite{englert4}. 

These days, many of us who have not yet attained the wisdom that 
comes with an {\it \'emeritat}, but who share Fran{\c c}ois' 
enthusiasm for all of physics' mysteries, are participants in 
the hunt for a still elusive theory, called ``M Theory'', which 
is to unify all known consistent string theories and to relate 
them through a web of non-perturbative dualities \cite{M,T}. 
This theory would also accommodate $d=11$ supergravity \cite{CJS} 
as a strong coupling limit via the relations
$$
R_{11} = \ell_s g_s \qquad  \ell_P^3 = \ell_s^3 g_s
$$
where $\ell_P$ is the $d=11$ Planck length, $\ell_s$ the string length, 
$g_s$ the string coupling constant, and $R_{11}$ the radius of the 
circle on which $d=11$ supergravity is compactified to ten dimensions
(the limit is taken in such a way that $\ell_P$ stays finite
while $g_s\rightarrow\infty$ and $\ell_s\rightarrow 0$, hence
$R_{11}\rightarrow\infty$). It is clear from these relations 
that present knowledge covers only the ``boundary'' of M Theory 
(where either the massive string modes or the $d=11$ Kaluza Klein 
modes are sent to infinity), but tells us almost nothing about 
its ``bulk'' --- the true domain of quantum gravity. Still, we 
can probably anticipate that it will be a pregeometrical 
theory in the sense that space time as we know it will emerge 
as a derived concept and that it should possess a huge symmetry 
group involving new types of Lie algebras (such as hyperbolic 
Kac Moody algebras) and perhaps even more general structures 
such as quantum groups.

According to the currently most popular proposal, M Theory ``is'' 
the $N\rightarrow\infty$ limit of the maximally supersymmetric 
quantum mechanical $SU(N)$ matrix model (see e.g. \cite{dW} for 
reviews and many references). This model had already 
appeared in an earlier study of the $d=11$ supermembrane 
in a flat background in the light cone gauge, and for any 
finite $N$, it can alternatively be obtained by dimensional 
reduction of the maximally extended supersymmetric Yang Mills 
theory in $d=10$ with gauge group $SU(N)$ to one (time) dimension. 
However, while matrix theory is pregeometrical in the sense that 
the target space coordinates are replaced by matrices, thus implying 
a kind of non-commutative geometry, the symmetries of
dimensionally reduced supergravities that we are concerned with
here, are hard to come by.

In this contribution, I will briefly describe some recent work 
done in collaboration with S.~Melosch \cite{NM}, and with 
K. Koepsell and H. Samtleben \cite{KNS2}, which was
motivated by recent advances in string theory (as well as the 
possible existence of an Ashtekar-type canonical formulation 
of $d=11$ supergravity). Although, at first sight, this work may
seem to have little to do with the issues raised above, 
it could actually be relevant in the context of M Theory, 
assuming (as we do) that further progress will crucially 
depend on the identification of the underlying symmetries, 
and that the hidden exceptional symmetries of maximal supergravity 
theories discovered long ago \cite{CJ,Julia1} may provide important 
clues as to where we should be looking. Support for this strategy 
derives from the fact that some local symmetries 
of the dimensionally reduced theories can be ``lifted" 
back to eleven dimensions. More precisely, it was 
shown in \cite{dewnic1,nic1} that there exist new versions of $d=11$ 
supergravity with local $\su$ and $\so$ tangent space symmetry, 
respectively. In both versions the supersymmetry variations 
acquire a polynomial form from which the corresponding formulas 
for the maximal supergravities in four and three dimensions can 
be read off directly and without the need for complicated duality 
redefinitions. This reformulation can thus be regarded as a step 
towards the complete fusion of the bosonic degrees of freedom of $d=11$ 
supergravity (i.e. the elfbein and the antisymmetric tensor $A_{MNP}$)
in a way which is in harmony with the hidden symmetries 
of the dimensionally reduced theories.

The existence of alternative versions of $d=11$ supergravity, 
which, though equivalent on-shell to the original version of 
\cite{CJS}, differ from it off-shell, strongly suggests the 
existence of a novel kind of ``exceptional geometry'' for $d=11$ 
supergravity, and thus the bigger theory containing it. This new 
geometry would be intimately tied to the special properties of the 
exceptional groups, and would be characterized by relations 
which have no analog in ordinary Riemannian geometry. Much of the
ongoing work centers on the role of extended objects (such as
2- and 5-branes in eleven dimensions), which couple
to the antisymmetric tensor fields present in $d=11$
and $d=10$ supergravities. Since these antisymmetric tensors
are here ``dualized away'', our formulation might open new vistas 
on a unified description of the basic ``objects'' of M Theory.

We will here concentrate on the $\so$ invariant version of 
$d=11$ supergravity \cite{nic1}. To derive it from the original 
formulation of $d=11$ supergravity, one first breaks the original 
tangent space symmetry SO(1,10) to its subgroup $SO(1,2)\times SO(8)$  
through a partial choice of gauge for the elfbein, and subsequently 
enlarges it again to $SO(1,2) \times SO(16)$ by introducing 
new gauge degrees of freedom. The construction thus requires a 
3+8 split of the $d=11$ coordinates and indices, implying a similar 
split for all tensors of the theory. The symmetry enhancement 
of the transverse (helicity) group $SO(9) \subset  SO(1,10)$
to $SO(16)$ requires suitable redefinitions of the bosonic and 
fermionic fields, or, more succinctly, their combination into 
tensors w.r.t. the new tangent space symmetry. 
It is important, however, that the dependence on all eleven 
coordinates is retained throughout.

In the bosonic sector, the elfbein and the three-index photon
are combined into new objects covariant w.r.t. to $d=3$ coordinate
reparametrizations and the new tangent space symmetry $SO(1,2)\times SO(16)$
(similar redefinitions must be made for the fermionic fields, but
we will not give explicit formulas here for lack of space).
In a special Lorentz gauge the elfbein takes the form
$$
E_M^{~A} = \left(\begin{array}{cc} 
            \Delta^{-1}e_\mu^{~a} & B_\mu^{~m} e_m^{~a}\\
            0& e_m^{~a}   \end{array} \right)
\label{11bein}
$$
where curved $d=11$ indices are decomposed as $M=(\mu ,m)$ 
with $\mu =0,1,2$ and $m= 3,...,10$ (with a similar decomposition
of the flat indices), and $\Delta := {\rm det} \, e_m^{~a}$.
It thus contains the (Weyl rescaled) dreibein and the Kaluza Klein 
vectors $B_\mu{}^{m}$, all of which are left untouched. By contrast,
we will trade the internal achtbein $e_m^{~a}$ for a rectangular 248-bein 
$e^m_\AC \equiv (e^{m}_{IJ},e^{m}_{A})$ containing the remaining 
``matter-like'' degrees of freedom, where the index $\AC\equiv ([IJ],A)$ 
labels the 248-dimensional adjoint representation of $E_{8(8)}$ in 
the $SO(16)$ decomposition ${\bf 248} \rightarrow {\bf 120}\oplus {\bf 128}$.
This 248-bein, which in the reduction to three dimensions contains 
all the propagating bosonic matter degrees of freedom of $d=3,N=16$ 
supergravity, is defined in a special $SO(16)$ gauge by
$$
(e^m_{IJ},e^m_A ) := \left\{ \begin{array}{ll}
     \Delta^{-1} e_a^{~m} \Gamma^a_{\alpha \dot \beta} 
            & \mbox{if $[IJ]$ or $A = (\alpha \dot \beta)$}\\
       0 & \mbox{otherwise}
       \end{array} \right.
$$
where the $SO(16)$ indices $IJ$ or $A$ are decomposed w.r.t. the diagonal 
subgroup $ SO(8)\equiv (SO(8)\times SO(8))_{diag}$ of $SO(16)$ 
(see \cite{nic1} for details). Being the inverse densitized 
internal achtbein contracted with an SO(8) $\Gamma$-matrix, 
this object is similar to the inverse densitized triad in 
Ashtekar's reformulation of Einstein's theory \cite{A}.

In addition  we need composite fields 
$\QC_\mu^\AC \equiv (Q_{\mu}^{IJ}, P_{\mu}^{A})$ 
and $\QC_m^\AC\equiv(Q_{m}^{IJ}, P_{m}^{A})$, which together 
make up an $E_{8(8)}$ connection again {\em in eleven dimensions}. 
Their explicit expressions in terms of the $d=11$ coefficients 
of anholonomity and the four-index field strength $F_{MNPQ}$ are, 
however, too lengthy to reproduce here \cite{nic1}.

The new geometry is encoded into constraints between 
the vielbein components, which rely in an essential way on 
special properties of the exceptional group $E_{8(8)}$. With the 
$E_{8(8)}$ indices ${\cal A},{\cal B},\dots (=1,\dots,248)$, we have
$$
({\cal P}_j)_{\cal A B}{}^{\cal C D} e^m_{\cal C} e^n_{\cal D} = 0
$$
where ${\cal P}_j$ are the projectors onto the $j = {\bf 1}\, ,\,
{\bf 248}$ and $\bf 3875$ representations of $E_{8(8)}$.
(Note that the projectors onto the $j= {\bf 27000}$ and
$\bf 30380$ representations do {\em not} vanish.) In addition, 
the 248-bein and the new connection fields are subject to 
a ``vielbein postulate" similar to the usual vielbein postulate, 
which states the covariant constancy of the 248-bein w.r.t. 
to an $E_{8(8)}$ covariant derivative involving the $E_{8(8)}$ 
connection $\QC_M^\AC$. For instance, for $M=m$ we have
$$
\partial_m e^n_\AC  + f_{\AC\BC}{}^\CC \QC_m^\BC e^n_\CC = 0
$$
where $f^{\AC\BC\CC}$ are the $E_{8(8)}$ structure constants.
(The relations with $M=\mu$ involve the Kaluza Klein
vectors $B_\mu{}^m$ and are slightly more complicated).
The supersymmetry variations of $d=11$ supergravity can now be 
re-expressed entirely in terms of these new variables and 
their fermionic partners \cite{nic1,NM}. 

Despite the ``$E_{8(8)}$ covariance'' of these relations, 
it must be stressed, however, that the full theory does 
not respect $E_{8(8)}$ invariance, as is already obvious from 
the fact that the fermions do not fit into representations 
of $E_{8(8)}$. However, the algebraic relations given above 
can be exploited to show \cite{KNS2} that there exists 
an $E_{8(8)}$ matrix $\VC$ {\em in eleven dimensions} 
such that
$$
e^m_\AC = \frac1{60} {\rm Tr} \Big( Z^m \VC X_\AC \VC^{-1}\Big)
$$
where the $X_\AC$ are the generators of $E_{8(8)}$, and the 
$Z^m$ span an eight-dimensional nilpotent subalgebra of $E_{8(8)}$
(there are altogether $36=8+28$ such nilpotent generators, whose role 
in relating the various dualized forms of dimensionally reduced 
supergravity has been explained in \cite{CJPL}). Because the 
fundamental and the adjoint representations of $E_{8(8)}$
are the same, we have $\VC X_\AC \VC^{-1} = X_\BC \VC^\BC{}_\AC$
and can thus rewrite this relation in the form
$$
e^m_\AC = \VC^m{}_\AC
$$
This means that the (inverse densitized) achtbein, which itself
is part of the elfbein of $d=11$ supergravity, has become part 
of an $E_{8(8)}$ matrix $\VC$ in eleven dimensions! Furthermore, 
it then follows from the generalized vielbein postulate stated above
that the $M=m$ part of the $E_{8(8)}$ connection $\QC_M^\AC$ can
be simply expressed in terms of this matrix via
$$
\QC_m = \VC^{-1} \partial_m \VC
$$
This simple formula, however, does not work for the low dimensional
components $\QC_\mu^\AC$.

The results obtained so far suggest further extensions
incorporating infinite dimensional symmetries. More specifically, 
the fact that the construction outlined above works with a 
4+7 and 3+8 split of the indices suggests that we should be 
able to construct versions of $d=11$ supergravity with infinite 
dimensional tangent space symmetries, which would be based 
on a 2+9 or even a 1+10 split of the indices. This would also 
be desirable in view of the fact that the new versions are 
``simple'' only in their internal sectors, as put in evidence
by the above formula for $\QC_m^\AC$. The general strategy 
would thus be to further enlarge the internal sector 
by absorbing more and more degrees of freedom into it, such that 
in the final step, only an einbein would be left in the low 
dimensional sector. However, it  is also clear that the elaboration 
of these ideas will not be an easy task. After all, it took a 
considerable effort extending over many years to show that the 
general pattern continues when one descends to $d=2$ and that 
the hidden symmetries become infinite dimensional, generalizing 
the Geroch group of general relativity \cite{Geroch}. 

There is some reason to believe that a generalization along these 
lines will take us beyond $d=11$ supergravity. The fundamental object 
of the theory could then turn out to be an infinite generalization
of the vierbein of general relativity, which would be acted 
upon from one side by a vast extension of the Lorentz group, 
containing not only space-time, but also internal symmetries, 
and perhaps even local supersymmetries. For the left action, 
one would have to appeal to some kind of generalized covariance 
principle, which would involve 
the $E_{11-d}$ symmetries.

To put these ideas into perspective, let us recall some facts 
about dimensionally reduced maximal supergravity to two dimensions. 
Following the empirical rules of dimensional reduction one is led 
to predict $E_9 = E_8^{(1)}$ as a symmetry for the dimensional reduction 
of $d=11$ supergravity to two dimensions \cite{Julia1,Julia2}.
This expectation is borne out by the existence of a linear 
system for maximal $N=16$ supergravity in two dimensions 
\cite{nic2} (see \cite{BM} for the bosonic theory, and \cite{JN1}
for a more recent summary). As is usually the case for integrable
systems, the linear system requires the introduction of an extra
spectral parameter $t$, and the extension of the $\sigma$-model 
matrix $\VC (x)$ to a matrix $\vh(x;t)$ depending on this extra 
parameter $t$. An unusual feature is that, due to the presence of 
gravitational degrees of freedom, this parameter becomes 
coordinate dependent, i.e. we have $t=t(x;w)$, where $w$ is 
an integration constant, sometimes referred to as the ``constant 
spectral parameter'' whereas $t$ itself is called the 
``variable spectral parameter''.

The (finite dimensional) coset structure of the higher dimensional 
theories has a natural continuation in two dimensions, with the only 
difference that the symmetry groups are infinite dimensional. This 
property is manifest from the transformation properties of the linear 
system matrix $\vh$, with a global affine symmetry acting from the left,
and a local symmetry corresponding to some ``maximal compact'' 
subgroup acting from the right:
$$
\vh (x;t) \longrightarrow g(w) \vh(x;t) h(x;t)
$$
Here $g(w)\in E_{9(9)}$ with affine parameter $w$, and the subgroup to 
which $h(x;t)$ belongs is defined as follows \cite{Julia2,BM,JN1}.
Let $\tau$ be the involution characterizing the coset space 
$E_{8(8)}/SO(16)$: then $h(t)\in\sxx$ is defined to consist 
of all $\ti$ invariant elements of $E_{9(9)}$, where the extended 
involution $\ti$ is defined by $\ti(h(t)):= \tau h(\ve t^{-1})$, 
with $\ve=+1$ (or $-1$) for a Lorentzian (Euclidean) worldsheet. 
Observe that $\sxx$ is different from the affine extension 
of $\s$ for either choice of sign.

Introducing a suitable triangular gauge and taking into account the
compensating $\sxx$ transformations to re-establish the chosen gauge
where necessary, one finds that these symmetries are realized in a 
non-linear and non-local fashion on the basic physical fields.
Moreover, they act as duality transformations in the sense that 
they mix scalar fields with their duals. At the linear level, 
a scalar field $\vf$ and its dual $\tilde\vf$ in two dimensions
are related by
$$
\partial_\mu \tilde\vf = \epsilon_{\mu\nu} \partial^\nu \vf
$$ 
If we were just dealing with free fields (as in conformal field
theory), there would not be much more to duality than this simple
equation, since a second dualization obviously brings us back 
to the original field (up to an integration constant). The crucial
difference here is that, as a consequence of the non-linearity 
of the field equations, there are {\em infinitely many} dual 
potentials because each dualization now produces a new 
(i.e. higher order) dual potential. It is basically this 
non-linearity inherited from Einstein's equations which explains 
why the group of duality transformations becomes infinite 
dimensional in two dimensions. Remarkably, however, already the 
free field relation above (with $\vf$ replaced by any target space 
coordinate) is central to modern string duality --- for instance
implying the emergence of D(irichlet) branes through the interchange 
of Neumann and Dirichlet boundary conditions for open strings \cite{Pol}.
It is furthermore well known that the integration constant arising 
in the dualization of a compactified string target space coordinate 
is associated with string winding modes, and that duality interchanges
Kaluza Klein and winding modes. Since we here get infinitely 
many such integration constants (i.e. one for every dualization),
we are led to predict the existence of an infinite tower of 
novel ``winding modes'' over and above the ones seen so far 
seen in string theory. These could be related to the mysterious 
states found in \cite{EGKR} that cannot be accounted for by the
standard counting arguments.

By representing the ``moduli space of solutions'' $\MC$ of the 
bosonic equations of motion of $d=11$ supergravity with nine 
commuting space-like Killing vectors as
$$
\MC = \frac{{\rm solutions \, of \, field \, equations}}%
             {{\rm diffeomorphisms}} 
    = \frac{E_{9(9)}}{\sxx} \label{coset1}
$$
one has managed to endow this space, which a priori is very 
complicated, with a group theoretic structure that makes it 
much easier to handle. In particular, the integrability of the 
system is directly linked to the fact that $\MC$ possesses an 
infinite dimensional ``isometry group'' $E_{9(9)}$. The introduction
of infinitely many gauge degrees of freedom embodied in the
subgroup $\sxx$ linearizes and localizes the action of this
isometry group on the space of solutions. Of course, 
in making such statements, one should keep in mind that a
mathematically rigorous construction of such spaces is a 
thorny problem. We can ignore these subleties here, not least 
because these spaces ultimately will have to be ``quantized'' anyway.

Elevating the local symmetries of maximal supergravity in two 
dimensions to eleven dimensions would thus require the existence 
of yet another extension of the theory, for which the Lorentz 
group $SO(1,10)$ is replaced by $\sx$ (the subgroup $\sxx$ can 
be interpreted as an extension of the transverse group $SO(9)$ 
in eleven dimensions). Accordingly, we would now decompose 
the elfbein into a zweibein and nine Kaluza Klein vectors 
$B_\mu^{~m}$ (with $m=2,...,10$). The remaining internal 
neunbein would have to be replaced by an ``Unendlichbein'' 
(or ``$\infty$-bein'', for short) $e^m_{\AC}(x;t)$. 
The parameter $t$ is necessary in order to 
parametrize the infinite dimensional extension of the symmetry 
group; whether it would still be a ``spectral parameter'' in the 
conventional sense of the word for the ``lifted'' theory, remains 
to be seen. One important difference with the dimensionally 
reduced theory is, however, clear: in eleven dimensions, 
there is no anolog of the dualization mechanism, which 
would ensure that despite the existence of infinitely many 
dual potentials, there are only finitely many physical degrees 
of freedom. This means that the construction will almost certainly 
take us beyond $d=11$ supergravity. 

Some information can be deduced from the requirement that in 
the dimensional reduction to $d=2$, there should exist a formula 
relating $e^m_\AC (x;t)$ to the linear system matrix $\vh(x;t)$, 
analogous to the one relating $e^m_\AC (x)$ to the $E_{8(8)}$ matrix 
$\VC (x)$ before. For this purpose, we would need a ninth nilpotent 
generator to complement the $Z^m$'s; an obvious candidate is the central
charge generator $c$, since it obeys $\langle c|c \rangle =
\langle c| Z^m \rangle = 0$ for all $m=3,...,10$. The parameter 
$t$, introduced somewhat ad hoc for the parametrization of the 
$\infty$-bein, must coincide in the dimensional reduction 
with the spectral parameter of the $d=2$ theory. Furthermore, 
the generalized ``$\infty$-bein postulate'' should reduce 
to the linear system of $d=2$ supergravity in this reduction. 

One difference with the previous situation, where the tangent 
space symmetry was still finite, is that the Lie algebra 
of $\sxx$ also involves the non-compact $E_{8(8)}$ generators, 
but in such a way that the generalized Cartan Killing form
on $E_{9(9)}$ is still positive on all these generators. This
follows from consideration of the $t$ dependence of the linear 
system of the dimensionally reduced theory and shows
that the new connections would constitute an $\sxx$ rather 
than an $E_{9(9)}$ gauge connection. This means that 
the covariantizations in the generalized vielbein postulate  
would be in precise correpondence with the local symmetries, 
in contrast with the previous relations which looked $E_{8(8)}$ 
covariant, whereas the full theory was actually invariant only 
under $\s$. Another curious feature is the following: in two 
dimensions, the linear system matrix contains all degrees 
of freedom, including the fermionic ones, and the local 
$N=16$ supersymmetry can be bosonized into a local
$\sxx$ gauge transformation \cite{NW}. This could mean
that there is a bosonization of fermions in the sense that
$e^m_\AC(x;t)$ would describe bosonic and fermionic 
degress of freedom. 

What has been said here could be summarized as follows: in
searching for a possible candidate M Theory, one should not
only concentrate on dimensionally reduced maximally extended 
{\em rigidly} supersymmetric theories (= supersymmetric Yang Mills
theories), but also consider the dimensionally reduced maximally 
extended {\em locally} supersymmetric theory. The idea (already
proposed in \cite{nic2}) is that a third quantized version of
maximal supergravity in two dimensions would give rise 
via a kind of bootstrap to a theory beyond $d=11$ supergravity 
that would contain the latter in the same way as superstring 
theories contain $d=10$ supergravity and $d=10$ 
super-Yang-Mills theories as special limits. However, it is not 
clear how (and if) this idea fits with presently accepted 
points of view.

\vspace{.75cm}


\noindent {\bf Acknowledgements}: I am very grateful to
the organizers for inviting me to this splendid event.
I would also like to thank K.~Koepsell, S.~Melosch and 
H.A.J.~Samtleben for the enjoyable collaboration on which
the new and as yet unpublished results reported here are based, 
and B.~de Wit for discussions.

\vspace{.75cm}



\begin{thebibliography}{99}

\bibitem{englert1} R.~Brout, F.~Englert, E.~Gunzig,
{\em Gen. Rel. Grav.} {\bf 10} (1979) 1
\bibitem{englert2} F.~Englert, H.~Nicolai, CERN-TH-3711 (1983)
\bibitem{englert3} B.~Biran, B.~de Wit, F.~Englert, H.~Nicolai,
{\em Phys. Lett.} {\bf 124B} (1983) 45
\bibitem{englert4} F.~Englert, H.~Nicolai, B.~Schellekens,
    {\em Nucl. Phys.} {\bf B274} (1986) 315
\bibitem{M} E.~Witten, {\em Nucl. Phys.} {\bf B443} (1995) 85,
  {\em Nucl. Phys.} {\bf B460} (1995) 335
\bibitem{T}
P.K.~Townsend, {\em Phys. Lett.} {\bf B350} (1995) 184, hep-th/9612121
\bibitem{CJS} E.~Cremmer, B.~Julia, J.~Scherk,
              {\em Phys. Lett.} {\bf 76B} (1978) 409
\bibitem{dW} B. de Wit, hep-th/9701169; \\
   T.~Banks, hep-th/9710231
\bibitem{NM} S.~Melosch, H.~Nicolai, {\em Phys. Lett.} {\bf B416} (1998) 91
\bibitem{KNS2} K.~Koepsell, H.~Nicolai, H.A.J.~Samtleben, 
              {\it work in progress}
\bibitem{CJ} E.~Cremmer, B.~Julia, {\em Nucl. Phys.} {\bf B159} (1979) 141
\bibitem{Julia1} B.~Julia, in {\em Superspace and Supergravity},
  eds. S.W.~Hawking and M.~Rocek (Cambridge University Press, 1981) 
\bibitem{dewnic1} B.~de Wit, H.~Nicolai, 
       {\em Nucl. Phys.} {\bf B274} (1986) 363
\bibitem{nic1} H.~Nicolai, {\em Phys.Lett.} {\bf 187B} (1987) 363 
\bibitem{A} A.~Ashtekar, {\em Phys. Rev. Lett.} {\bf 57} (1986) 2244 
\bibitem{CJPL}E.~Cremmer, B.~Julia, C.N.~Pope, H.~Lu,
   {\em Nucl. Phys.} {\bf B523} (1998) 73
\bibitem{Geroch} R.~Geroch, {\em J. Math. Phys.} {\bf 13} (1972) 394;\\
W.~Kinnersley, D.M.~Chitre, {\em J. Math. Phys.} {\bf 18} (1977) 1538
\bibitem{Julia2} B.~Julia, in {\em Unified theories and beyond}, 
Proc. 5th Johns Hopkins Workshop on Current Problems in Particle
Theory, Johns Hopkins University, Baltimore, 1982
\bibitem{nic2} H.~Nicolai, {\em Phys. Lett.} {\bf 194B} (1987) 402
\bibitem{BM} P.~Breitenlohner, D.~Maison, {\em Ann. Inst. H.~Poincar\'e}
    {\bf 46} (1987) 215
\bibitem{JN1} B.~Julia, H.~Nicolai, {\em Nucl. Phys.} {\bf B482} (1996) 431
\bibitem{Pol} J.~Polchinski, hep-th/9611050
\bibitem{EGKR} S.~Elitzur, A.~Giveon, D.~Kutasov, E.~Rabinovici,
     hep-th/9707217
\bibitem{NW} H.~Nicolai, N.P.~Warner, 
  {\em Comm. Math. Phys.} {\bf 125} (1989) 384

\end{thebibliography}
\end{document}